\documentclass[options]{JHEP3}

\usepackage{graphicx}% Include figure files
\title{Some Impacts of Lorentz Violation on Cosmology}
\author{Arianto$^{(1,2)}$, Freddy P. Zen$^{(1)}$, Bobby E. Gunara$^{(1)}$, Triyanta$^{(1)}$ and Supardi$^{(1,3)}$\\

$^{(1)}$Theoretical Physics Laboratory, THEPI,\\ Faculty of Mathematics and Natural Sciences,
 Institut Teknologi Bandung \\
Jl. Ganesha 10 Bandung 40132, Indonesia.\\

$^{(2)}$Department of Physics, Udayana University \\
Jl. Kampus Bukit Jimbaran Kuta-Bali 80361, Indonesia. \\

$^{(3)}$Department of Physics, Sriwijaya University \\
Jl. Raya Palembang-Prabumulih, Inderalaya, Indonesia. \\}

\abstract{ The impact of Lorentz violation on the dynamics of a scalar field is investigated. In particular, we study the dynamics of a scalar field in the scalar-vector-tensor theory where the vector field is constrained to be unity and time like. By taking a generic form of the scalar field action, a generalized dynamical equation for the scalar-vector-tensor theory of gravity is obtained to describe the cosmological solutions. We present a class of exact solutions for an ordinary scalar field or phantom field corresponding to a power law coupling vector and the Hubble
parameter. As the results, we find a constant equation of state in de Sitter space-time and power law expansion with the quadratic of coupling vector, while a dynamic equation of state is obtained for $n> 2$. Then, we consider the inflationary scenario based on the Lorentz violating scalar-vector-tensor theory of gravity with general power-law coupling vector and two typical potentials: inverse power-law and power-law potentials. In fact, both the coupling vector and the potential models affect the dynamics of the inflationary solutions. Finally, we use the dynamical system formalism to study the attractor behavior of a cosmological model containing a scalar field endowed with a quadratic coupling vector and a chaotic potential.}


\begin{document}

%===============================================================%
%************************ SECTION I ****************************%
%===============================================================%
\section{Introduction}
Scalar field theory has become the generic playground for building cosmological models related to particle physics, in particular for obtaining inflationary cosmologies which is one of the most reliable concepts to describe the early stage of the universe. The key property of the laws of physics that makes inflation possible is the existence of states of matter that have a high energy density which cannot be rapidly lowered. The Inflationary scenario~\cite{Lindeetc} relies on the potential energy of a scalar (inflaton) field to drive a period of early universe acceleration. It has been thought that the early universe could be well characterized by a series of phase transitions, in which topological defects could be formed~\cite{Vilenkin}. In the context of the string theory, the natural values of the gauge and gravitational couplings in our 4d universe are explained by the dynamics of 'moduli' scalar fields~\cite{4, 5}. Moreover, originating from the work of Sen~\cite{Sen} (see also \cite{Gibbons}), the possibility of the tachyon field being a candidate for the inflaton has been extensively studied. The tachyon action is of the Dirac Born Infeld form~\cite{Ashoke Sen} which leads to an equation of state interpolating between $-1$ at early times and $0$ at late-times. This suggests the possibility that the tachyon can play the role of the inflaton at early times and the dark matter at late-times. Serious difficulties, however, plagues the tachyonic inflation~\cite{Kofman}. These include large density perturbations, problem with reheating and formation of caustics.

Recent observational evidences especially from the Type Ia Supernovae \cite{Riess, Jassal} and WMAP satellite missions \cite{Bennett:2003bz}, indicate that we live in a favored spatially flat universe consisting approximately of 30\% dark matter and 70\% dark energy. In the framework of the General Relativity this means that about two thirds of the total energy density of the Universe consists of dark energy, the still unknown component with a relativistic negative pressure $p < -\rho/3$. The simplest candidate for dark energy is the cosmological $\Lambda$-term. During the cosmological evolution the $\Lambda$-term component has the constant (Lorentz invariant) energy density $\rho$ and
pressure $p = -\rho$. However, it has got the famous and serious fine-tuning problem, while the also elusive dark matter candidate might be a lightest and neutral supersymmetry particle with only gravity interaction. For this reason the different forms of dynamically changing dark energy with an effective equation of state $w < -1/3$ were proposed, instead of the constant vacuum energy density. As a particular example of dark energy, the scalar field with a slow rolling potential (quintessence) \cite{Wetterich} is often considered. The possible generalization of quintessence is a k-essence \cite{Picon}, the scalar field with a non-canonical Lagrangian. Any such behavior would have
far-reaching implications for particle physics. However, recent theory of gravity with the Lorentz violation ~\cite{Sato:2000vu,Coleman:1998ti} are proposed. 

More recently, authors in Ref.~\cite{KS} explored the Lorentz violating scenario in the context of the scalar-vector-tensor theory. They showed that the Lorentz violating vector affects the dynamics of the inflationary model. One of the interesting features of this scenario, is that the exact Lorentz violating inflationary solutions are related to the absence of the inflaton potential. In this case, the inflation is completely associated with the Lorentz violation. Depending on the value of the coupling parameter, the three kind of exact solutions are found: the power law inflation, de Sitter inflation, and the super-inflation. 

The purpose of this paper is to study the dynamics of a scalar field in the framework of Lorentz violating scalar-vector-tensor model, taking into account the effect of the dynamically coupling vector. In this framework, we explore a class of exact solutions such as evolution of a scalar field and equation of state parameters. We discuss an inflationary scenario with a power-law coupling vector model with two typical potentials: an inverse power-law potential and a power-law potential. Then, we show that it is possible to find attractor solutions in the Lorentz violating scalar-vector-tensor model in which both the coupling function and the potential function are specified.

The organization of this paper is the following: in Section~\ref{secII}, we set down the general formalism for the
scalar-vector-tensor theory where the Lorentz symmetry is spontaneously broken due to the unit-norm vector field. In
Section~\ref{secIII}, we use our formalism to find an exact solution of the equation of state. In Section~\ref{secIV}, we study Lorentz violating (inverse) power law inflation. In Section~\ref{secV}, the critical points of the global system and their stability are presented. The final Section is devoted to the conclusions.

%===============================================================%
%************************ SECTION II ****************************%
%===============================================================%
\section{ General Formalism}\label{secII}

In the present section, we develop the general reconstruction
scheme for the scalar-vector-tensor theory. We will consider the
properties of general four-dimensional universe, i.e. the universe
where the four-dimensional space-time is allowed to contain any
non-gravitational degree of freedom in the framework of Lorentz
violating scalar-tensor-vector theory of gravity. Let us assume
that the expectation values of a vector field $u^\mu$
is $<0| u^\mu u_\mu |0> = -1$. The action can be written as the
sum of three distinct parts:
\begin{eqnarray}
     S&=& S_g + S_u + S_{\phi} \ ,
     \label{eq:action}
\end{eqnarray}
where the actions for the tensor field $S_g$, the vector field
$S_u$, and the scalar field $S_{\phi}$ are,
\begin{eqnarray}
        S_g &=& \int d^4 x \sqrt{-g}~ {1\over 16\pi G}R  \ ,
    \label{eq:act-grav} \\
  S_u &=& \int d^4 x \sqrt{-g} \left[
  - \beta_1 \nabla^\mu u^\nu \nabla_\mu u_\nu
   -\beta_2 \nabla^\mu u^\nu \nabla_\nu u_\mu    -\beta_3 \left( \nabla_\mu u^\mu \right)^2 \right. \nonumber\\
 && \left.
  -\beta_4 u^\mu u^\nu \nabla_\mu u^\alpha \nabla_\nu u_\alpha
    + \lambda \left( u^\mu u_\mu +1 \right) \right]  \ ,
  \label{eq:act-VT} \\
  S_{\phi}  &=&  \int d^4 x \sqrt{-g}~ {\cal{L}}_{\phi}  \ .
   \label{eq:act-matter}
\end{eqnarray}
In the above $\beta_i(\phi)$ ($i=1,2,3,4$) are arbitrary parameters which has the dimension of mass squared. It means that $\sqrt{\beta_i}$ gives the mass scale of symmetry breakdown. 
${\cal{L}}_{\phi}$ is the Lagrangian density for scalar field,
expressed as a function of the metric $g_{\mu\nu}$ and the scalar
field $\phi$. Then, the action~(\ref{eq:action}) describes the
scalar-tensor-vector theory of gravity. The dimensionless vector field, $u^\mu$, satisfies the constraint
\begin{eqnarray}
  u^\mu u_\mu = -1 .
\end{eqnarray}

For the background solutions, we use the homogeneity and isotropy
of the universe  spacetime
\begin{eqnarray}
ds^2 = - {\mathcal{N}}^2 (t) dt^2 + e^{2\alpha(t)} \delta_{ij} dx^i dx^j \ ,
\end{eqnarray}
where ${\mathcal{N}}$ is a lapse function and the scale of the universe is determined by $\alpha$. We take the constraint
\begin{eqnarray}
   u^\mu = \left( {1\over {\mathcal{N}}} , 0 ,0 ,0 \right) \ ,
\end{eqnarray}
where ${\mathcal{N}} =1$ is taken into account after the variation. Varying the action (\ref{eq:action}) with respect to $g^{\mu\nu}$, we have field equations
\begin{eqnarray}
   R_{\mu\nu}-{1\over 2}g_{\mu\nu}R = 8\pi G T_{\mu\nu} \ ,
   \label{eq:einstein-eq}
\end{eqnarray}
where $T_{\mu\nu} =T_{\mu\nu}^{(u)} + T_{\mu\nu}^{(\phi)}$ is the total energy-momentum tensor,
$T_{\mu\nu}^{(u)}$ and $T_{\mu\nu}^{(\phi)}$ are the energy-momentum tensors of vector
and scalar fields, respectively, defined by the usual formulae
\begin{eqnarray}
     T_{\mu\nu}^{(k)} = -2\frac{\partial {\cal{L}}^{(k)} }{\partial g^{\mu\nu}
     }+ g_{\mu\nu} {\cal{L}}^{(k)}  , \qquad k=u, {\phi}  \ .
\end{eqnarray}
The time and space components of the total energy-momentum tensor
are given by
\begin{eqnarray}
     T^{0}_{0} = - \rho_u -\rho_{\phi} \ , \qquad    T^{i}_i =  p_u+ p_{\phi} \ ,
     \label{eq:00-ii-compot}
\end{eqnarray}
where the energy density and pressure of the vector field are
given by
\begin{eqnarray}
    && \rho_u =  -3\beta H^2  \ ,
    \label{eq:rho-vf} \\
    &&p_u =  \left(3 + 2{H^{\prime}\over H} + 2{\beta^{\prime}\over \beta} \right)\beta H^2 \ ,
    \label{eq:pres-vf} \\
    && \beta \equiv \beta_1 +3 \beta_2 + \beta_3 \ .
    \label{eq:def-beta}
\end{eqnarray}
From the above equations, one can see that $\beta_4$ does not contribute to the background
dynamics. A prime denotes the derivative of any quantities $X$ with
respect to $\alpha$. $X^{\prime}$ is then related to its
derivative with respect to $t$ by
$X^{\prime}=(dX/dt)H^{-1}=\dot{X} H^{-1}$ where
$H=d\alpha/dt=\dot{\alpha}$ is the Hubble parameter. From Eqs.
(\ref{eq:rho-vf}) and (\ref{eq:pres-vf}), one obtains the
energy equation for the vector field $u$
\begin{eqnarray}
   {\rho}^{\prime}_u + 3({\rho}_u + p_u)=+3H^2 \beta^{\prime}  \ ,
   \label{eq:eos-vec}
\end{eqnarray}
and for the scalar field
\begin{eqnarray}
    {\rho}^{\prime}_{\phi} + 3({\rho}_{\phi} + p_{\phi})=-3H^2 \beta^{\prime}  \ .
   \label{eq:eos-mat}
\end{eqnarray}
The total energy equation in the presence of both the
vector and the scalar fields is, accordingly,
\begin{equation}
   {\rho}^{\prime} + 3({\rho} + p)=0 \ , \quad (\rho = \rho_u + \rho_{\phi}) \ .
   \label{eq:eos-total}
\end{equation}
This energy conservation equation can also be obtained by
equating the covariant divergence of the total energy-momentum
tensor to zero, since the covariant divergence of the Einstein
tensor is zero by its geometric construction. It follows from
contraction of the geometric Bianchi identity.

Substituting Eq.~(\ref{eq:00-ii-compot}) into the Einstein
equations (\ref{eq:einstein-eq}), we obtain two independent
equations, called the Friedmann equations, as follows:
\begin{eqnarray}
     && -3H^2 = 8\pi G \left(3\beta H^2 -\rho_{\phi} \right)  \ ,
    \label{eq:Friedmann1} \\
    && -2HH' - 3H^2 =  8\pi G \left[\left(3 + 2{H^{\prime}\over H} + 2{\beta^{\prime}\over \beta} \right)\beta H^2 +p_{\phi} \right] \ .
     \label{eq:Friedmann2}
\end{eqnarray}
These Friedmann equations can be rewritten as
\begin{eqnarray}
    && \left( 1 + \frac{1}{8\pi G \beta} \right) H^2={1\over 3\beta} \rho_{\phi}   \ ,
    \label{eq:Friedmann} \\
    && \left( 1 + \frac{1}{8\pi G \beta} \right) \left( HH'+H^2\right)=-{1\over 6} \left( {\rho_{\phi}\over \beta} + {3p_{\phi}\over \beta} \right) - H^2 {\beta'\over \beta} \ .
    \label{eq:Friedmannsec}
\end{eqnarray}
The second term on RHS of Eq.~(\ref{eq:Friedmannsec}) is a
consequence of the coupling vector field as a function of scalar
field. If $\beta_i =0$, thus without the vector field, the above
equations reduce to the conventional ones. And in the case
$\beta=const$., the above equations are lead to the Friedmann equations given in Ref.
\cite{Carroll:2004ai}.

Using Eqs.~(\ref{eq:Friedmann}) and (\ref{eq:eos-mat}), we obtain a set of equations as follows:
\begin{eqnarray}
&&  {H^{\prime}\over H} + {\bar{\beta}^{\prime}\over \bar{\beta}}
+ {3\over 2}(1+\omega_{\phi}) =0 \ ,
\label{eq:hb}  \\
&&  {H^{\prime}\over H} - {\rho^{\prime}_{\phi}\over \rho_{\phi}}
- {3\over 2}(1+\omega_{\phi})=0 \ ,
\label{eq:hr}  \\
&&  {\rho^{\prime}_{\phi}\over \rho_{\phi}} +
{\bar{\beta}^{\prime}\over \bar{\beta}} + 3(1+\omega_{\phi})=0
\label{eq:rb}  \ ,
\end{eqnarray}
where
\begin{eqnarray}
   \bar{\beta} &=&   \beta + {1\over 8\pi G} \ ,
   \label{eq:twopart}
\end{eqnarray}
and $\omega_{\phi} ={p_{\phi} /\rho_{\phi}}$ is the equation of
state of the scalar field. It is easy to check that the equations (\ref{eq:hb})--(\ref{eq:rb}) satisfy the following
constraint
\begin{eqnarray}
&&  2{H^{\prime}\over H} + {\bar{\beta}^{\prime}\over \bar{\beta}}
- {\rho^{\prime}_{\phi}\over \rho_{\phi}}=0    \ .
\label{eq:cls}
\end{eqnarray}
In order to solve the Eqs.~(\ref{eq:hb})--(\ref{eq:rb}) and
(\ref{eq:cls}), we have to specify the model and the matter
content of the universe. The general solution of these equations
can be written as
\begin{eqnarray}
&&  H\bar{\beta} \propto \exp \left[-\int {3\over
2}(1+\omega_{\phi}(\alpha)) d\alpha \right] \ ,
\label{eq:hb-sol}  \\
&&  {H\over \rho_{\phi}} \propto \exp \left[\int {3\over
2}(1+\omega_{\phi}(\alpha)) d\alpha \right] \ ,
\label{eq:hr-sol}  \\
&&  \rho_{\phi} \bar{\beta} \propto   \exp \left[-\int
3(1+\omega_{\phi}(\alpha)) d\alpha \right] \label{eq:rb-sol}  \ .
\end{eqnarray}
If the functions $\omega_{\phi}$ and $\bar{\beta}$ are given,
then we can find the evolution of the Hubble parameter under the
Lorentz violation. For example, the cosmological constant
corresponds to a fluid with a constant equation of state
$\omega_{\phi}=-1$. Thus the above equations reduce to: $
H\bar{\beta} \propto 1$, $H \propto \rho_{\phi}$ and
$\rho_{\phi}\bar{\beta} \propto 1$ where $H$, $\rho_{\phi}$ and
$\beta$ are functions of $\alpha$. If $\omega_{\phi}$ is a constant parameter of a simple one component fluid, and for a given $\alpha(t)$, Eqs. (\ref{eq:hb})--(\ref{eq:rb}) can be used to
determine $\beta(\alpha)$ and $\rho_{\phi}(\alpha)$. We, then, are able to determine the potential of the Lorentz violation model.

%===============================================================%
%************************ SECTION III **************************%
%===============================================================%
\section{Dynamical Equations for Scalar Fields}\label{secIII}
For a given scalar field Lagrangian with the FRW background, we
can obtain the equations of motion for a scalar field by using
Eq.~(\ref{eq:eos-mat}) and Eqs.~(\ref{eq:hb})--(\ref{eq:rb}). Let
us consider the Lagrangian density of a scalar field $\phi$ with a
potential $V(\phi)$ in Eq.~(\ref{eq:action}):
\begin{eqnarray}
{\cal{L}}_{\phi}= -{\eta\over 2}(\nabla \phi)^2 - V(\phi) \ ,
\end{eqnarray}
where $(\nabla \phi)^2=g^{\mu\nu}\partial_{\mu}\phi\partial_{\nu}\phi$. Ordinary scalar fields correspond to $\eta = 1$ while $\eta = -1$ is for phantoms. For the homogeneous field the density
$\rho_{\phi}$ and pressure $p_{\phi}$ of the scalar field,
 may be found as follows
\begin{eqnarray}
&&\rho_{\phi} = {\eta\over 2} H^2 \phi^{\prime 2} + V(\phi) \ ,
\label{eq:infrho} \\
&&p_{\phi}= {\eta\over 2} H^2 \phi^{\prime 2} - V(\phi) \ .
\label{eq:infp}
\end{eqnarray}
The corresponding equation of state parameter is, accordingly
\begin{eqnarray}
\omega_{\phi}={p_{\phi}\over \rho_{\phi}} = - \frac{1- \eta H^2
\phi^{\prime 2}/2V}{1 + \eta H^2 \phi^{\prime 2}/2V} \ .
\label{phant-eos}
\end{eqnarray}
Substituting Eq.~(\ref{eq:infrho}) into Eq.~(\ref{eq:Friedmann}), the
Friedmann equation leads to
\begin{eqnarray}
H^2  = \frac{1}{3\bar{\beta}} \left[
  \frac{\eta}{2} H^2 \phi^{\prime 2} + V(\phi) \right] \ .
  \label{eq:01-1}
\end{eqnarray}
Now, differentiating Eq.~(\ref{eq:infrho}) with respect to
$\alpha$ and using Eq.~(\ref{eq:eos-mat}), and also differentiating Eq.~(\ref{eq:01-1}) with respect to
$\alpha$ and using Eq.~(\ref{eq:03-1}) give, respectively,
\begin{eqnarray}
  \phi'' &=&- \left( \frac{H'}{H} + 3\right)\phi' - \eta \frac{V_{,\phi}}{H^2}
             - 3\eta \bar{\beta}_{,\phi} \ ,
             \label{eq:03-1}\\
\phi^{\prime} &=&-2\eta \bar{\beta}\left(\frac{H_{,\phi}}{H} +
\frac{\bar{\beta}_{,\phi}}{\bar{\beta}} \right) \ .
     \label{eq:02-1}
\end{eqnarray}
Substituting Eq.~(\ref{eq:02-1}) into the Friedmann equation the
potential of the scalar field can be written as
\begin{eqnarray}
    V = 3\bar{\beta} H^2 \left[ 1-{2\over 3}\eta\bar{\beta}\left({\bar{\beta}_{,\phi}\over \bar{\beta}} + {H_{,\phi}\over H}\right)^2 \right] \ .
\end{eqnarray}
Note that in the above equations the Hubble parameter $H$ has been
expressed as a function of $\phi$, $H=H(\phi(t))$. From
Eq.~(\ref{eq:hb}), the equation of state can be written as
\begin{eqnarray}
    \omega_\phi &=& -1 + {4\over 3}\eta\bar{\beta}\left(\frac{H_{,\phi}}{H} + \frac{\bar{\beta}_{,\phi}}{\bar{\beta}} \right)^2 \nonumber\\
    &=&-1 + {1\over 3}\eta \frac{\phi^{\prime 2}}{\bar{\beta}} \ .
    \label{eos1-1}
\end{eqnarray}
Equations (\ref{eq:02-1}) and (\ref{eos1-1}) are two equations
that we need to solve for the scalar field $\phi$ and the equation
of state $\omega_\phi$. This is achieved only if the Hubble parameter
$H(\phi)$ and the coupling vector ${\bar{\beta}}(\phi)$ are known.
For different choice of the Hubble parameter $H(\phi)$ and the
coupling vector ${\bar{\beta}}(\phi)$, it is possible to extract a
class of exact solutions of Eqs.~(\ref{eq:02-1}) and
(\ref{eos1-1}). We shall solve Eqs.~(\ref{eq:02-1}) and
(\ref{eos1-1}) to obtain the following physical quantities ($V$ and $K$ are the
potential and  kinetic energies, respectively):
\begin{eqnarray}
   &&V={3\over 2}(1-\omega_\phi){\bar{\beta}}H^2, \quad K={3\over 2}(1+\omega_\phi){\bar{\beta}}H^2, \nonumber\\
   &&\rho_\phi=3{\bar{\beta}}H^2 \ ,~~\quad\quad \qquad p_\phi=3\omega_\phi {\bar{\beta}}H^2 \ .
   \label{quant-phys}
\end{eqnarray}

In the following two subsections we will explore a class of exact
solutions.

\subsection{Exact solutions and the behavior of scalar fields}
We shall have to solve equations (\ref{eq:02-1}) and
(\ref{eos1-1}) for $H$, $\omega_\phi$, $\bar{\beta}$, and $V$,
which is not possible unless two are known. In the present
subsection, we consider an example to find an exact solution of
the equation of state of the scalar field in the quadratic
coupling vector. The equation of state for the scalar field has
been intensively studied in \cite{Zlatev} for the so called
tracking cosmological solutions introduced in \cite{Steinhardt99},
and some classes of potentials allowing for the field equation of
state were described.

Let us consider a simple model
\begin{equation}
    H=H_0 \ , \qquad  \bar{\beta}(\phi) = m\phi^2 \ ,
    \label{model-hconst1}
\end{equation}
where $H_0$ and $m$ are positive constant parameters. The equation
(\ref{eq:02-1}) can now be integrated to yield the evolution of
the scalar field
\begin{equation}
   \phi(t)=\phi_0 \exp \left[-4\eta mH_0(t-t_0) \right] \ ,
   \label{sol-n2}
\end{equation}
where $\phi(t=t_0)\equiv \phi_0$ is a constant. Then, it is easy
to find the equation of state of the scalar field by using
Eq.~(\ref{eos1-1}). We obtain
\begin{eqnarray}
   \omega_\phi &=& -1 + {16\over 3} m \ , \quad \textit{for ordinary scalar fields} \ ,
   \label{const-eos-ord} \\
    \omega_\phi &=& -1 - {16\over 3} m \ , \quad  \textit{for phantom fields} \ .
   \label{const-eos-phant}
\end{eqnarray}
Then, the potential and the kinetic energies, the energy density and the
pressure of the scalar field evolve according to
\begin{eqnarray}
   V(t) &=&  mH_0^2\phi_0^2(3 - 8m)\exp \left[-8\eta mH_0(t-t_0) \right]  \ , \\
    K(t) &=&  8\eta\left(mH_0\phi_0\right)^2 \exp \left[-8\eta mH_0(t-t_0) \right]\ , \\
    \rho(t) &=& 3mH_0^2\phi_0^2 \exp \left[-8\eta mH_0(t-t_0) \right] \ , \\
    p(t) &=& \eta m H_0^2\phi_0^2 \left( 16m -3\eta \right)\exp \left[-8\eta mH_0(t-t_0) \right] \ .
\end{eqnarray}

The above solutions are completely associated with the Lorentz
violation. The model (\ref{model-hconst1}) depicts that the cosmic
evolution starts from a constant value of the scale factor and
grows exponentially, $a(t)=a_0 e^{H_0(t-t_0)}$. The coupling vector
decreases exponentially for the
ordinary scalar field and increases exponentially for the phantom field from a constant value of $m\phi_0^2$. Hence
the potential energy, the kinetic energy, the energy density and the pressure decrease
exponentially for the ordinary scalar field. For the
phantom field, on the other hand, the potential energy and energy density as well as the absolute values of the kinetic energy and the pressure increase exponentially. Note that the
kinetic energy and the pressure begin with the negative values. The Eqs.~(\ref{const-eos-ord})--(\ref{const-eos-phant}) show that the equation of state $\omega_\phi$ is non-dynamical because it only
depends on the value of the coupling vector parameter $m$, both for
the ordinary scalar and the phantom fields. Since an
accelerated expansion occurs for $\omega_\phi < -1/3$ then we
have $m<1/8$ for the ordinary scalar field. However, the present
data of the Universe seems to tell that $\omega_\phi$ might be
less than $-1$. Thus, the value $m$ may be chosen in order to fit
the present observable constraint on the equation of state
parameter.

In other case, for instance, $H(\phi)=H_0\phi^\xi$ and
$\bar{\beta}(\phi)=m\phi^2$, we also find the constant equation of
state,
\begin{eqnarray}
   \omega_\phi = -1 + {4\over 3} \eta m (\xi + 2)^2 \ .
   \label{const-eos-gen}
\end{eqnarray}
The condition for the accelerating Universe $\ddot{a}$ or $H'/H >
-1$ yields
\begin{eqnarray}
   \eta m < {1\over 2\xi (\xi + 2)} \ .
   \label{cond-acc}
\end{eqnarray}
This model gives a power law expansion
\begin{eqnarray}
    {a(t)\over a_o}= \left[1 + {H_0\phi_0^\xi \over p}(t-t_0) \right]^p \ , \qquad p>1 \ ,
\end{eqnarray}
where
\begin{eqnarray}
     p={1\over 2\eta m\xi(\xi+2)} \ .
\end{eqnarray}
The scalar field evolve as
\begin{equation}
   \phi(t)=\phi_0\left(1+{H_0\phi_0^\xi\over p}(t-t_0)\right)^{-1/\xi} \ .
   \label{phi-lain}
\end{equation}
Hence, the complete set of solutions is found by substituting
Eqs.~(\ref{const-eos-gen}) and (\ref{phi-lain}) into
Eqs.~(\ref{quant-phys}).

In the following subsection, we will see that the equation of
state may be dynamics. For this purpose we generalize the coupling
vector to $\bar{\beta}(\phi)=m\phi^n$, $n>2$.

\subsection{Variable equation of states}
Let us consider a model where the coupling vector is a power law
of the scalar field,
\begin{equation}
    H=H_0 \ , \quad  \bar{\beta}(\phi) = m\phi^n \ ,\quad n >2 \ ,
    \label{model-hconst}
\end{equation}
where $H_0$, $m$ and $n$ are constant positive parameters. Following the
same above procedure, the scalar $\phi$ can be evaluated as,
\begin{equation}
   \phi(t)=\frac{\phi_0}{ \left[1 + 2\eta mnH_0(n-2)\phi_0^{n-2}(t-t_0) \right]^{1\over{n-2}}} \ ,
\end{equation}
the coupling vector is given by
\begin{equation}
   \bar{\beta}(t)=\frac{m\phi_0^n}{ \left[1 + 2\eta mnH_0(n-2)\phi_0^{n-2}(t-t_0) \right]^{n\over{n-2}}} \ ,
\end{equation}
and the dynamical equation of state (\ref{eos1-1}) is
\begin{eqnarray}
  \omega(t) = -1 + \frac{4\eta mn^2\phi_0^{n-2}/3}{1+2\eta mnH_0(n-2)\phi_0^{n-2}(t-t_0)} \ .
  \label{eos-narbri}
\end{eqnarray}
Then, the potential and kinetic energies, the energy density and the pressure of the
scalar field are given by
\begin{eqnarray}
   V(t) &=&  3mH_0^2\phi_0^n\left[1- \frac{2\eta mn^2\phi_0^{n-2}/3}{1+2\eta mnH_0(n-2)\phi_0^{n-2}(t-t_0)}\right] \times \nonumber\\
    &&\times {1\over\left[1 + 2\eta mnH_0(n-2)\phi_0^{n-2}(t-t_0) \right]^{{n\over{n-2}}}} \ ,\\
    K(t) &=& \frac{2\eta\left(mnH_0\phi_0^{n-1}\right)^2}{ \left[1 + 2\eta mnH_0(n-2)\phi_0^{n-2}(t-t_0) \right]^{{2(n-1)\over{n-2}}}} \ , \\
    \rho(t) &=& \frac{3mH_0^2\phi_0^n}{ \left[1 + 2\eta mnH_0(n-2)\phi_0^{n-2}(t-t_0) \right]^{{n\over{n-2}}}} \ , \\
    p(t) &=& 3mH_0^2\phi_0^n \left[ -1 + \frac{4\eta mn^2\phi_0^{n-2}/3}{1+2\eta mnH_0(n-2)\phi_0^{n-2}(t-t_0)} \right] \nonumber\\
    &&\times {1\over\left[1 + 2\eta mnH_0(n-2)\phi_0^{n-2}(t-t_0) \right]^{{n\over{n-2}}}} \ .
\end{eqnarray}

Thus, the model (\ref{model-hconst}) describes that the cosmic
evolution grows exponentially from a constant value of the scale factor, $a(t)=a_0 e^{H_0(t-t_0)}$, while the coupling
vector $\bar{\beta}$ started from a constant value of the scalar
field, $m\phi_0^n$. The equation of state $\omega_\phi$ is
dynamical both for the ordinary scalar and phantom fields. Then the
potential energy, kinetic energy, the energy density and the pressure decrease
for the ordinary scalar field. For the phantom
field, on the other hand,  the potential and energy density increase while the kinetic
energy and pressure begin with the negative values.
%===============================================================%
%************************ SECTION IV ***************************%
%===============================================================%
\section{ Lorentz Violating Inflation Scenario}\label{secIV}
As it has been studied by authors in Ref.~\cite{KS}, the Lorentz
violation on the inflationary scenario can be divided into two
parts: the Lorentz violations stage $8\pi G \beta \gg 1$ and the
standard slow roll stage $8\pi G \beta \ll 1$. The first stage
corresponds to $\bar{\beta} = \beta$ in Eq.~(\ref{eq:twopart}) and
the second stage corresponds to $\bar{\beta} = 1/ 8\pi G $, then
we have the usual dynamical equations.

In this section we will consider the inflationary scenario for the scalar field (inflaton). In particular, we consider a power-law coupling vector, $\beta(\phi) = m \phi^n$, with two types of the potential: $V(\phi) = \mu^{4+\nu}\phi^{-\nu}$ and $V(\phi) = {1\over 2}M^2\phi^2$. Here $\mu$, $\nu$ and $M$ are parameters. Thus, the dynamics of each particular inflationary model are determined by the Friedmann equation and the scalar field equation of motion once the functional form of the inflaton potential and the coupling parameter have been specified. Let us collect the dynamics-related equations for the inflaton the Friedmann equation (\ref{eq:01-1}) in inflationary models
\begin{eqnarray}
 H^2  = \frac{1}{3\bar{\beta}} \left[
  \frac{1}{2} H^2 \phi^{\prime 2} + V(\phi) \right] \ ,
  \label{eq:01}
\end{eqnarray}
the constraint equation (obtained from Eqs.~(\ref{phant-eos})  and (\ref{eq:hb}))
\begin{eqnarray}
\frac{H'}{H} + \frac{1}{2} \frac{\phi^{\prime 2}}{\bar{\beta}} + \frac{\bar{\beta}'}{\bar{\beta}} =0 \ ,
     \label{eq:02}
\end{eqnarray}
and  the equation of motion Eq.~(\ref{eq:03-1})
\begin{eqnarray}
\phi'' + \frac{H'}{H} \phi' + 3\phi' + \frac{V_{,\phi}}{H^2}+ 3 \bar{\beta}_{,\phi} =0  \ .
             \label{eq:03}
\end{eqnarray}
$\bar{\beta}$ is given by Eq.~(\ref{eq:twopart}). Then, at the critical value of $\phi$, the effective coupling vector becomes
\begin{eqnarray}
8\pi G \beta (\phi_c ) =1  \ .
\end{eqnarray}
For example, a coupling parameter of the form $\beta=m\phi^2$ gives the critical value
\begin{eqnarray}
   \phi_c = \frac{M_{pl}}{\sqrt{8m \pi }}  \ ,        M_{pl} = G^{-1} \ .
\end{eqnarray}
Let $\phi_i$ be the corresponding initial value of the scalar
field. Putting $\phi_i \sim 3 M_{pl}$, the
Lorentz violation implies the criterion $m > 1/(72\pi) \sim 1/226$.

The set of Eqs.~(\ref{eq:01})--(\ref{eq:03}) constitutes the
equations we have to solve for the problem specified by
the coupling parameter $\beta(\phi)$
 and the potential $V (\phi)$. In following subsection, we  consider with a model with  the coupling
parameter $\beta(\phi)$ is given by
\begin{eqnarray}
  \beta(\phi) = m \phi^n \ ,
  \label{eq:model-1}
\end{eqnarray}
where $n$ and $m$ are parameters. For the model
(\ref{eq:model-1}), we obtain the critical value of the scalar
field and the criterion for Lorentz violation
\begin{eqnarray}
  \phi_c = \left( \frac{M_{pl}^2}{8m \pi }\right)^{1/n} \quad \textit{and}\quad m > \frac{M_{pl}^2}{8\pi(3M_{pl})^{n} } \ .
\end{eqnarray}

Now, we consider two typical potentials appear in many cosmological implications: an inverse power-law
potential and  a power-law potential. We discuss those solutions
and analyze the two regimes separately.

%===============================================================%
%************************ SUBSECTION IV ************************%
%===============================================================%
\subsection{Inverse power law potential: $V(\phi) = \mu^{4+\nu}\phi^{-\nu}$}
In this subsection, we consider the class of power law potential
\begin{eqnarray}
  V(\phi) = \mu^{4+\nu}\phi^{-\nu} \ ,
\end{eqnarray}
where $\mu$ and $\nu$ are constants. Inverse power law models are
interesting for a number of reasons. In conventional cosmology,
they drive 'intermediate' inflation~\cite{Barrow} and typically
produce significant tensor perturbations for almost
scale-invariant scalar fluctuations. They arise in supersymmetric
condensate models of QCD~\cite{Binetruy} and can in principle act
as a source of quintessence~\cite{Wetterich, Balbi}.

%===============================================================%
%************************ SUBSUBSECTION ************************%
%===============================================================%
\subsubsection{Lorentz violating stage}
Let us first consider the Lorentz violating stage, $8\pi G \beta \gg 1$ ($\bar{\beta}=\beta$), we
have the equations (\ref{eq:01})--(\ref{eq:03}). In this stage both the coupling function and the potential function are relevant.

An inflationary epoch, in which the scale factors $a$ are
accelerating, requires the scalar field $\phi$ to evolve slowly
compared to the expansion of the universe. Thus, the following
conditions of slow-rolling are required:
\begin{eqnarray}
 H^2 \phi^{\prime 2} \ll V \ , \quad \phi'' \ll \phi' \ , \quad \phi^{\prime 2} \ll \beta , \quad \textit{and} \quad \beta' \ll \beta \ .
\label{eq:slowroll-inflaton}
\end{eqnarray}
The formalism which gives these slow roll conditions are discussed
in Ref. \cite{KS}. This is sufficient to guarantee inflation.
Under the slow-roll conditions Eq.~(\ref{eq:slowroll-inflaton}),
the Eqs.~(\ref{eq:01})--(\ref{eq:03}) can be simplified.
We obtain the slow roll equations
\begin{eqnarray}
   H^2 \simeq \frac{V}{3\beta} \ , \quad \textit{and} \quad \phi' \simeq - \beta\left({\beta_{,\phi} \over \beta}+ \frac{V_{,\phi}}{V} \right)    \ .
   \label{eq:sl-inflaton-1}
\end{eqnarray}
Inserting Eq.~(\ref{eq:model-1}) and the potential of the form
$V(\phi) = \mu^{4+\nu}\phi^{-\nu}$ into
Eq.~(\ref{eq:sl-inflaton-1}), we have
\begin{eqnarray}
   H^2 &=& \frac{\mu^{4+\nu}}{3m} \phi^{-(\nu+n)} \ ,
   \label{eq:eolut-sol-Frid-inflaton}\\
   \phi' &=& -m(n-\nu)\phi^{(n-1)} \ .
   \label{eq:eolut-sol-scl-inflaton}
\end{eqnarray}
One can then solve for $\phi$ from
Eq.~(\ref{eq:eolut-sol-scl-inflaton}),
\begin{eqnarray}
   \phi(\alpha) = \left[\phi_i^{2-n}+m(n-2)(n-\nu)(\alpha-\alpha_i)\right]^{-{1\over n-2}}  \ , \quad \textit{for} \quad n \neq 2, n \neq \nu \ ,
   \label{eq:eolut-sol1-scl-inflaton}
\end{eqnarray}
where $\phi(\alpha =\alpha_i)\equiv \phi_i$ is a constant. The
Friedmann equation gives
\begin{eqnarray}
   H^2(\alpha) = \frac{\mu^{4+\nu}}{3m} \left[\phi_i^{2-n}+m(n-2)(n-\nu)(\alpha-\alpha_i)\right]^{{n+\nu \over n-2}} \ .
   \label{eq:eolut-sol1-Frid-inflaton}
\end{eqnarray}
The solution (\ref{eq:eolut-sol1-scl-inflaton}) and the slow roll conditions (\ref{eq:slowroll-inflaton}) during the
Lorentz violating stage give $n>2$ because
\begin{eqnarray}
&& \phi^{\prime 2}\left(\sim \alpha^{-2(1-n)/(2-n)}\right) \ll \beta\left(\sim \alpha^{n/(2-n)}\right) \ , \\
&&\beta' \left(\sim \alpha^{-2(1-n)/(2-n)}\right) \ll \beta
\left(\sim \alpha^{n/(2-n)}\right) \ .
\end{eqnarray}
From Eq.~(\ref{eq:eolut-sol1-Frid-inflaton}), the universe
expands during the Lorentz violating stage as
\begin{eqnarray}
   {a(t) \over a_i} = \exp \left\{ -{B\over C}+ {1\over C} \left( {1\over B^{D-1}}-AC(D-1)(t-t_i) \right)^{-{1\over D-1}} \right\} \ , \quad C \neq 0 \ ,
   \label{eq:scale-univ-inflaton}
\end{eqnarray}
where the constants $A, B, C$ and $D$ are
\begin{eqnarray}
   A=\sqrt{{\mu^{4+\nu}\over 3m}} \ , \quad B=\phi_i^{2-n} \ , \quad C=m(n-2)(n-\nu) \ , \quad D=\frac{n+\nu}{2(n-2)} \ .
\end{eqnarray}
Combining Eqs. (\ref{eq:scale-univ-inflaton}),
(\ref{eq:eolut-sol1-scl-inflaton}) and
(\ref{eq:eolut-sol1-Frid-inflaton}), we obtain the
physical quantities
\begin{eqnarray}
  && \phi(t) = \left( {1\over B^{D-1}}-AC(D-1)(t-t_i) \right)^{{1\over (D-1)(n-2)}}  \ ,
   \label{eq:eolut-sol1-scl-time} \\
  && H(t) = A \left( {1\over B^{D-1}}- AC(D-1)(t-t_i) \right)^{-{D\over (D-1)}} \ ,
   \label{eq:eolut-sol1-Frid-time}\\
  && \beta(t)=n\left( {1\over B^{D-1}}- AC(D-1)(t-t_i) \right)^{{n\over (D-1)(n-2)}} \ .
   \label{eq:eolut-sol1-beta-time}
\end{eqnarray}
The scalar field energy density, on the other hand, evolves according to
\begin{eqnarray}
   \rho (t)\simeq V =3mA^2\left( {1\over B^{D-1}}-AC(D-1)(t-t_i) \right)^{{n-2D(n-2)\over (D-1)(n-2)}}  \ .
\end{eqnarray}
One can see that the Hubble parameter $H$ decreases during the
Lorentz violation stage.

For $n=2, \nu\neq 2$, Eq.~(\ref{eq:model-1})
and the second part of Eq.~(\ref{eq:sl-inflaton-1}) gives
\begin{eqnarray}
   \phi (\alpha)= \phi_i e^{-n(2-\nu)(\alpha -\alpha_i)} \ ,
   \label{eq:sol1-sl-inflaton}
\end{eqnarray}
where $\phi(\alpha=\alpha_i)\equiv \phi_i$. For this solution to
satisfy slow roll conditions (\ref{eq:slowroll-inflaton}), we need
$ m <1/(2-\nu)^2$. Thus, we have the range $1/226 <m <1/(2-\nu)^2$
of the parameter for which the Lorentz violating inflation is
relevant. The Hubble parameter as a function of the scale factor,
$\alpha$, is given by
\begin{eqnarray}
H^2(\alpha) =  H^2_i e^{-m(\nu^2-4)(\alpha - \alpha_i)}   \ ,
  \label{eq:sol2-sl-inflaton}
\end{eqnarray}
where
\begin{eqnarray}
H^2_i =\frac{\mu^{4+\nu}}{3 m \phi_i^{(2+\nu)}} \ ,
\end{eqnarray}
while the scale factor $a(t) = e^{\alpha}$ is of the form:
\begin{eqnarray}
          {a(t)\over a_i} = \left[{\phi (t) \over \phi_i}\right]^{{1\over m(\nu-2)}} \ .
\end{eqnarray}
Now we obtain the evolution of some physical quantities as follows
\begin{eqnarray}
   && {a(t)\over a_i} =\left[m(\nu^2-4)H^2_i(t-t_i)\right]^{1\over m(\nu^2-4)}  \ ,
   \label{eq:eolut-m2-a} \\
  && {\phi(t)\over\phi_i}  = \left[m(\nu^2-4)H^2_i(t-t_i)\right]^{1\over \nu + 2} \ ,
   \label{eq:eolut-m2-scl-time} \\
  && {H(t)\over H^2_i} = \sqrt{m(4-\nu^2)(t-t_i)} \ ,
   \label{eq:eolut-m2-Frid-time}\\
  && {\beta(t)\over m\phi^2_i}= \left[m(\nu^2-4)H^2_i(t-t_i)\right]^{2\over \nu + 2} \ ,
   \label{eq:eolut-m2-beta-time}\\
   &&\rho (t)={\mu^{4+\nu}\over \phi_i}\left[m(\nu^2-4)H^2_i(t-t_i)\right]^{-{\nu \over \nu + 2}}  \ .
\end{eqnarray}
%

%===============================================================%
%************************ SUBSUBSECTION ************************%
%===============================================================%
\subsubsection{Standard slow roll stage}
The governing equations (\ref{eq:01})--(\ref{eq:03}) in the standard slow roll stage $8\pi G\beta \ll 1$ ($\bar{\beta}=(8\pi G)^{-1}$), are, accordingly,
\begin{eqnarray}
&&   H^2  = \frac{8\pi G}{3} \left[  \frac{1}{2} H^2 \phi^{\prime
2} + V  \right] \ ,
\label{eq:ssrs1}\\
 &&   \frac{H'}{H} + 4\pi G  \phi^{\prime 2}  =0  \ ,
 \label{eq:ssrs2}\\
&&  \phi'' + \frac{H'}{H} \phi' + 3\phi' + \frac{V_{,\phi}}{H^2}
=0 \ . \label{eq:ssrs3}
\end{eqnarray}
In this case the slow roll equations are given by
\begin{eqnarray}
 H^2  \simeq \frac{8\pi G}{3}  V   \ , \quad
\phi'\simeq - \frac{1}{8\pi G} \frac{V_{,\phi}}{V}       \ .
\end{eqnarray}
For the potential model $V(\phi) = \mu^{4+\nu}\phi^{-\nu}$, the
evolution of the inflaton and the Hubble parameter can be solved
as
\begin{eqnarray}
    \phi^2 (\alpha) &=&  \phi_c^2 + \frac{\nu}{4\pi G }(\alpha-\alpha_c) \ , \\
    H^2(\alpha)  &=& \frac{8\pi G}{3}   \mu^{4+\nu} \left[ \phi_c^2 + \frac{\nu}{4\pi G }(\alpha-\alpha_c)\right]^{-\nu/2} \ ,
\end{eqnarray}
and the scale factor is given by
\begin{eqnarray}
  {a (t)\over a_c}  = \exp \left[ {4\pi G \over \nu}( \phi^2 (t)  -\phi_c^2) \right] \ .
\end{eqnarray}
The evolution equations are given by
\begin{eqnarray}
   && {a(t)\over a_c} = \exp\left\{-{B_s\over C_s}+{1\over C_s} \left[B_s^{D_s+1} +A_s C_s(D_s+1)(t-t_c) \right]^{{1\over D_s+1}}\right\}  \ ,
   \label{eq:eolut-sdr-a} \\
  && \phi(t)= \left[B_s^{D_s+1} +A_s C_s(D_s+1)(t-t_c) \right]^{{1\over 2(D_s+1)}}  \ ,
   \label{eq:eolut-sdr-scl-time} \\
  && H(t) = A_s\left[B_s^{D_s+1} +A_s C_s(D_s+1)(t-t_c) \right]^{{D_s\over (D_s+1)}} \ ,
   \label{eq:eolut-sdr-Frid-time}
\end{eqnarray}
and the scalar field energy density evolves as
\begin{eqnarray}
  \rho(t)= {3\over 8}\left(\frac{A_s^2 C_s}{D_s} \right)\left[B_s^{D_s+1} +A_s C_s(D_s+1)(t-t_c) \right]^{{2D_s\over (D_s+1)}} \ .
   \label{eq:eolut-sdr-beta-time}
\end{eqnarray}
where $A_s$, $B_s$, $C_s$ and $D_s$ are the constants,
\begin{eqnarray}
  A_s = \sqrt{\frac{8\pi G}{3}   \mu^{4+\nu}} \ , \quad B_s =\phi_c^2 \ , \quad C_s =\frac{\nu}{4\pi G } \ , \quad D_s =\frac{\nu}{4} \ .
\end{eqnarray}
Note that, in the standard slow roll stage, the Hubble parameter
$H$ increases.

Another interesting quantity is the number of e-folding during the
inflationary phase. The total e-folding number reads
\begin{eqnarray}
  N &=& -{B\over C}+ {1\over C} \left( {1\over B^{D-1}}- AC(D-1)(t_c-t_i) \right)^{-{1\over D-1}} \nonumber\\
  &&-{B_s\over C_s}+{1\over C_s} \left[B_s^{D_s+1} +A_s C_s(D_s+1)(t_e-t_c) \right]^{{1\over D_s+1}} \nonumber\\
   &=&{1\over C} \left( \phi^{2-n}_c - \phi_i^{2-n} \right)+ \frac{4\pi G }{\nu} \left( \phi^2_e - \phi_c^2 \right) \ ,
\end{eqnarray}
for $m > 2, \nu \neq m$ and
\begin{eqnarray}
 N &=&{1\over n(\nu^2-4)}\log \left[n(\nu^2-4)H^2_i(t_c-t_i)\right] \nonumber\\
 && -{B_s\over C_s}+{1\over C_s} \left[B_s^{D_s+1} +A_s C_s(D_s+1)(t_e-t_c) \right]^{{1\over D_s+1}} \nonumber\\
  &=&{1\over m(2-\nu)} \log \left( {\phi_i \over \phi_c} \right)+ \frac{4\pi G }{\nu} \left( \phi^2_e - \phi_c^2 \right) \ ,
\end{eqnarray}
for $n = 2, \nu \neq 2$. Note that the first terms of the above
equations arise from the Lorentz violating stage. As an example,
let us take the values: $N=70$, $m = 10^{-2}$, $n=2$ and $\nu=1$.
If $\phi_e \sim 0.3 M_{pl}$ is the value of scalar field at the
end of inflation, then, $\phi_c \sim 2 M_{pl}$. The contribution
from the inflation end is still relevant. Therefore, we get
$\phi_i \sim 2.5 M_{pl}$.

%===============================================================%
%************************ SUBSECTION IV ************************%
%===============================================================%
\subsection{Power law potential: $V(\phi) = {1\over 2}M^2\phi^2$}
%===============================================================%
%************************ SUBSUBSECTION ************************%
%===============================================================%
\subsubsection{Lorentz violating stage}
The most realistic inflationary universe scenarios are chaotic
models. For the model $V(\phi) = {1\over 2}M^2\phi^2$, assuming
the slow roll conditions, we find the slow roll equations during
the Lorentz violating regime as follows
\begin{eqnarray}
   H^2 &=& \frac{M^2}{6n} \phi^{-(n-2)} \ ,
   \label{eq:eolut-sol-Frid-chao}\\
   \phi' &=& -m(n+2)\phi^{(n-1)} \ .
   \label{eq:eolut-sol-scl-chao}
\end{eqnarray}
Then we find the solution (\ref{eq:eolut-sol-scl-chao}) as
\begin{eqnarray}
   \phi(\alpha) = \left[\phi_i^{2-n}+m(n^2-4)(\alpha-\alpha_i)\right]^{{1\over 2-n}}  \ ,
   \label{eq:eolut-neq2-scl-inflaton}
\end{eqnarray}
for $m\neq 2$ and
\begin{eqnarray}
   \phi(\alpha) = \phi_i e^{-4m(\alpha-\alpha_i)}  \ ,
   \label{eq:eolut-e2-scl-inflaton}
\end{eqnarray}
for $n= 2$. The inflationary scenario of this model was already
obtained in Ref.~\cite{KS} where the Hubble parameter becomes
constant during the Lorentz violating regime and $1/226<m<1/16$ is
the range of parameter $m$. We concern here the solution for
$n\neq 2$. The solution for the Hubble parameter is given by
\begin{eqnarray}
   H^2(\alpha) = \frac{M^2}{6m} \left[\phi_i^{2-n}+m(n^2-4)(\alpha-\alpha_i)\right] \ ,
   \label{eq:eolut-sol1-Frid-chao}
\end{eqnarray}
and
\begin{eqnarray}
   {a(t)\over a_i}=  \exp \left[ \frac{1}{m(n^2-4)} \left( {1\over \phi^{n-2}(t)} -{1\over \phi_i^{n-2}}\right) \right] \ ,
   \label{eq:eolut-scale1-chao}
\end{eqnarray}
which is the solution for the scale factor. As in the previous
subsection, we also obtain $n>2$ which the effect of Lorentz
violation occurs in this regime. The time evolution of the above
equations can be obtained by integrating
Eq.~(\ref{eq:eolut-sol1-Frid-chao}), we get
\begin{eqnarray}
   \alpha(t) =\alpha_i  -{b\over c}+ {1\over c} \left[ b^{1/2}+{1\over 2}dc(t-t_i) \right]^2 \ ,
   \label{eq:eolut-scale1-time}
\end{eqnarray}
where
\begin{eqnarray}
   b=\phi_i^{2-n} \ , \quad c =m(n^2-4) \ , \quad d = \sqrt{\frac{M^2}{6m}} \ , \quad n > 2 \ .
\end{eqnarray}
Then the evolution equations are given by
\begin{eqnarray}
  && {a(t)\over a_i}=  \exp \left\{ -{b\over c}+{1\over c} \left[  b^{1/2} +{1\over2}d c(t-t_i) \right]^{2}\right\} \ ,
   \label{eq:1} \\
  && \phi(t) = \left[  b^{1/2} +{1\over2}d c(t-t_i) \right]^{-{2\over n-2}}  \ ,
   \label{eq:2} \\
 &&   H(t)= d \left[  b^{1/2} +{1\over2}d c(t-t_i) \right] \ ,
   \label{eq:3}\\
 && \beta(t) = n\left[  b^{1/2}+{1\over2}d c(t-t_i) \right]^{-{2n\over n-2}}  \ ,
 \label{eq:4}\\
  && \rho(t) = 3nd^2 \left[  b^{1/2} +{1\over2}d c(t-t_i) \right]^{-{4\over n-2}} \ .
 \label{eq:5}
\end{eqnarray}
Since $b$, $c$ and $d$ are positive constants, one can see
that the Hubble parameter $H$ and the scale factor $a$ increase
during the Lorentz violating stage for $n>2$. In the case $n=2$,
the Hubble parameter is constant. In the following subsection, we
will see that the Hubble parameter decreases in the standard slow
roll stage.

%===============================================================%
%************************ SUBSUBSECTION ************************%
%===============================================================%
\subsubsection{Standard slow roll stage}
Now, let us consider the chaotic inflationary scenario in the
standard slow roll stage. A set of the dynamical equations of the
scalar field are given by Eqs.~(\ref{eq:ssrs1})--(\ref{eq:ssrs3}).
Assuming the standard slow roll conditions, we find the slow roll
equations
\begin{eqnarray}
&&   H^2  \simeq \frac{4\pi G}{3}  M^2 \phi^2  \ , \\
&&  \phi' \simeq - \frac{1}{4\pi G} \phi^{-1}      \ .
\end{eqnarray}
The evolution of the inflaton can be solved as
\begin{eqnarray}
    \phi^2 (\alpha) = \phi_c^2 - \frac{1}{2\pi G }(\alpha-\alpha_c) \ .
\end{eqnarray}
The Hubble parameter and the scale factor $a(t) = e^{\alpha}$ can
be also obtained as
\begin{eqnarray}
    && H^2  = \frac{4\pi GM^2 }{3}  \left( \phi_c^2 - \frac{1}{2\pi G }(\alpha-\alpha_c) \right) \ ,
    \label{eq:h1} \\
    && a (t)  =a_c \exp \left[ 2\pi G ( \phi_c^2 -\phi^2 (t) ) \right] \ .
    \label{eq:a1}
\end{eqnarray}
From Eq.~(\ref{eq:h1}), we obtain
\begin{eqnarray}
    \alpha(t) -\alpha_c= {b_s\over c_s}-{1\over c_s} \left[b_s^{1/2} -{1\over 2}d_s c_s(t-t_c) \right]^{2} \ ,
\end{eqnarray}
and the dynamical evolutions are given by
\begin{eqnarray}
  && {a(t)\over a_c}=  \exp \left[  {b_s\over c_s}-{1\over c_s} \left(b_s^{1/2} -{1\over2}d_s c_s(t-t_c) \right)^{2}\right] \ ,
   \label{eq:1st} \\
  && \phi (t) = b_s^{1/2} -{1\over2}d_s c_s(t-t_c)  \ ,
   \label{eq:2nd} \\
 &&   H(t)  = d_s \left[b_s^{1/2} -{1\over2}d_s c_s(t-t_c) \right] \ ,
   \label{eq:3td}\\
  && \rho(t) = {3c_s d_s\over 4}\left[b_s^{1/2} -{1\over2}d_s c_s(t-t_c) \right]^{2} \ ,
 \label{eq:5fi}
\end{eqnarray}
with
\begin{eqnarray}
     b_s= \phi_c^2 \ , \quad c_s = \frac{1}{2\pi G } \ , \quad  d_s = \frac{4\pi GM^2 }{3} \ .
\end{eqnarray}
Note that the Hubble parameter decreases in the standard slow roll
stage.

In the case of chaotic potential, the total e-folding number reads
\begin{eqnarray}
    N &=& -{b\over c}+{1\over c} \left[  b^{1/2} +{1\over2}d c(t_c-t_i) \right]^{2} +{b_s\over c_s}-{1\over c_s} \left[b_s^{1/2} -{1\over2}d_s c_s(t_e-t_c) \right]^{2} \nonumber\\
    &=&\frac{1}{m(n^2-4)}\left( \phi_c^{2-n} - \phi_i^{2-n}\right) +2\pi G \left( \phi_c^2 -\phi_e^2 \right) \ ,
\end{eqnarray}
where $\phi_e$ is the value of scalar field at the end of
inflation. Notice that the first term arises from the Lorentz violating stage.

%===============================================================%
%************************ SECTION V ****************************%
%===============================================================%

\section{Phase-space analysis}\label{secV}
In this section, we investigate the global structure of the
dynamical system via phase plane analysis and compute the
cosmological evolution by numerical analysis. Introducing the
following variables:
\begin{eqnarray}
    &&x\equiv{\phi'\over \sqrt{6\bar{\beta}}} \ , \qquad y\equiv\sqrt{{V\over 3H^2\bar{\beta}}} \ ,
    \label{def-xy}\\
    && \lambda_1 \equiv -{\bar{\beta}_{,\phi}\over \sqrt{\bar{\beta}}} \ , \qquad \lambda_2 \equiv - \sqrt{\bar{\beta}}{V_{,\phi}\over V} \ ,
    \label{def-lambda}\\
    && \Gamma_1 \equiv \frac{\bar{\beta} \bar{\beta}_{,\phi\phi}}{\bar{\beta}_{,\phi}^2} \ , \qquad \Gamma_2 \equiv \frac{V V_{,\phi\phi}}{V_{,\phi}^2} \ ,
    \label{def-gamma}
\end{eqnarray}
the Eqs.~(\ref{eq:02}) and (\ref{eq:03}) can be written as a
plane-autonomous system
\begin{eqnarray}
    &&x'=-3x(1-x^2)+\sqrt{{3\over 2}} (\lambda_1+\lambda_2)y^2 \ ,
    \label{auto-x}\\
    && y' = \left[3x-\sqrt{{3\over 2}} (\lambda_1+\lambda_2)\right]xy \ ,
    \label{auto-y}\\
    && \lambda_1^{\prime} =-\sqrt{6}\lambda_1^2  \left(\Gamma_1 -{1\over 2}\right)x  \ ,
    \label{auto-L1}\\
    &&\lambda_2^{\prime} =-\sqrt{6}\lambda_2^2   \left[ \Gamma_2 -\left(1-{\lambda_1\over 2\lambda_2}\right) \right] x \ ,
    \label{auto-L2}
\end{eqnarray}
where the prime denotes a derivative with respect to the logarithm
of the scale factor, $\alpha=\ln a$. The functions
$\lambda_1(\phi)$ and $\lambda_2(\phi)$ determine a type of the
coupling vector and the potential, respectively. The Friedmann
constraint, Eq.~(\ref{eq:01}), takes the simple form
\begin{eqnarray}
    x^2 +y^2 =1 \ .
    \label{constrps}
\end{eqnarray}
The equation of state for the scalar field could be expressed in
terms of the new variables as
\begin{eqnarray}
    \omega_\phi =\frac{p_\phi}{\rho_\phi}=\frac{x^2-y^2}{x^2+y^2} \ .
\end{eqnarray}
Notice that $x^2$ measures the contribution to the expansion due
to the scalar field kinetic energy and the coupling function,
while $y^2$ measures the contribution to the expansion due to the
potential energy and the the coupling function.

Equations (\ref{auto-x})--(\ref{auto-L2}) are written as an
autonomous phase system of the form ${\bf x}'={\bf f}({\bf x})$
where ${\bf x}=(x,y,\lambda_1,\lambda_2)$. The use of this form
for the dynamical equations allows the fixed points of the system
to be readily identified, and the so-called critical points ${\bf
x}_0$ are solutions of the system of equations ${\bf f}({\bf
x}_0)=0$. To determine their stability we need to perform linear
perturbations around the critical points in the form ${\bf x} =
{\bf x}_0+{\bf u}$, which results in the following equations of
motion ${\bf u}' =M{\bf u}$, where
\begin{eqnarray}
   M_{ij} =\frac{\partial f_i}{\partial x_j}\Big |_{{\bf x}_0} \ .
\end{eqnarray}
In the case of the dynamical equations
(\ref{auto-x})--(\ref{auto-L2}), {\bf u} is a 4-column vector
consisting of the perturbations of $x$, $y$, $\lambda_1$ and
$\lambda_2$. Thus, $M_{ij}$ is a $4 \times 4$ matrix. The
stability of the critical points is determined by the eigenvalues
$\mu_i$ of the matrix $M$ at the critical points. A non-trivial
critical point is called stable (unstable) whenever the
eigenvalues  of $M$ are such that $Re(\mu_i) < 0$ ($Re(\mu_i) >
0$). If neither of the aforementioned cases are accomplished, the
critical point is called a saddle point.
\begin{figure}[http]% fig.1
\begin{center}
\includegraphics[height=10cm, width=12cm]{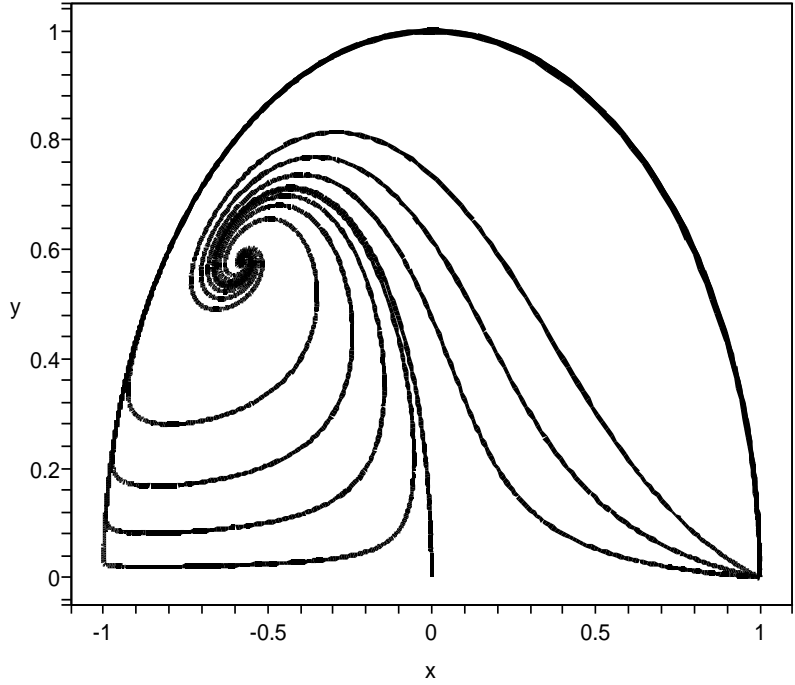}
\end{center}
\caption{The phase plane of Lorentz violating kinetic dominated
solution for $m>3/8$. }
\label{figure1}
\end{figure}

In the following, we will study the simplest model,
\begin{eqnarray}
    \bar{\beta}(\phi)= m \phi^2 \ , \qquad    V(\phi)=  {1\over 2}M^2  \phi^2 \ ,
     \label{model-ps}
\end{eqnarray}
where $m$ and $M$ are parameters. Substituting
Eqs.~(\ref{model-ps}) into Eqs.~(\ref{def-lambda}) and
(\ref{def-gamma}), respectively, we obtain
\begin{eqnarray}
    \lambda_1 = \lambda_2 = -2\sqrt{m} \ , \qquad   \Gamma_1=\Gamma_2={1\over 2} \ ,
\end{eqnarray}
and Eqs.~(\ref{auto-L1}) and (\ref{auto-L2}) are trivially
satisfied. In the former, Eqs.~(\ref{auto-x}) and (\ref{auto-y})
can be fused into the single equation,
\begin{eqnarray}
    x'&=&-\left[3x-\sqrt{{3\over 2}} (\lambda_1+\lambda_2)\right](1-x^2) \nonumber\\
        &=& -\left(3x+2\sqrt{6m}\right)(1-x^2) \ ,
    \label{auto-single}
\end{eqnarray}
which is one dimensional phase-space corresponding to the unit
circle. Critical points correspond to fixed points where ${\bf
x}'=0$, and there are Lorentz violation self-similar solutions
with
\begin{eqnarray}
    {H'\over H} &=& -3x^2 +\sqrt{6}\lambda_1 x \ .
    \label{self-sim}
\end{eqnarray}
Note that the second term arises from Lorentz violation. Applying
the above procedure, setting ${\bf x}'=0$, the critical points
$(x_0,y_0)$ of the system   are $(1,0)$, $(-1,0)$, and
$(-\sqrt{8m/3},\sqrt{1-8m/3})$. For any form of the potential in
the Lorentz violation stage, the critical points $(1,0)$ or
$(-1,0)$ correspond to two Lorentz violation kinetic-dominated
solutions. Then, the critical point
$(-\sqrt{8m/3},\sqrt{1-8m/3})$ corresponds to  a Lorentz violation
potential-kinetic solution. Integration of Eq.~(\ref{self-sim}) with respect to $\alpha$ will show
that all critical points, ${\bf x}_0$, correspond to the
Hubble parameter
\begin{eqnarray}
   H \propto \exp \left(-{\alpha\over p}\right)  \ .
\end{eqnarray}
\begin{figure}[http]% fig.1
\begin{center}
\includegraphics[height=10cm, width=12cm]{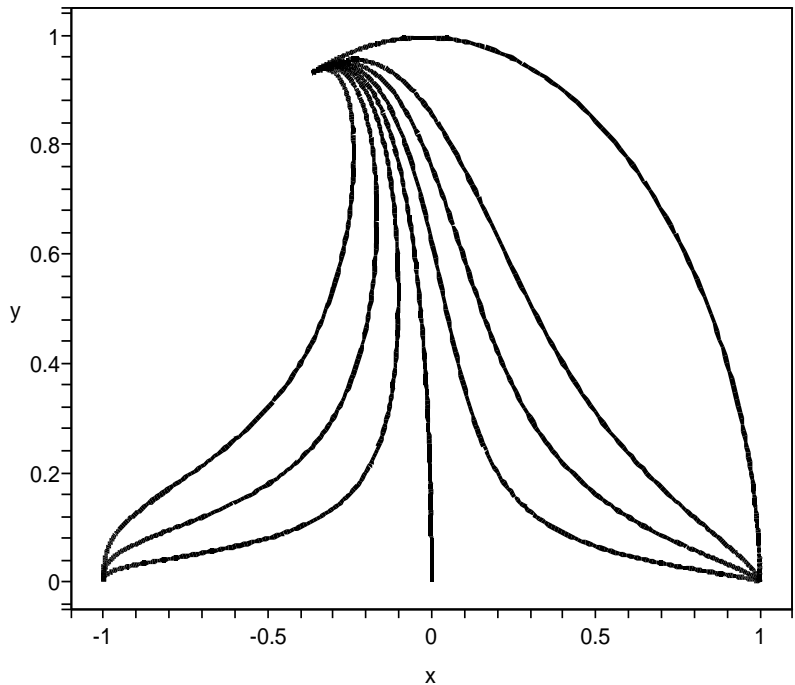}
\end{center}
\caption{The phase plane of Lorentz violating kinetic-potential
solution for $m<3/8$. } \label{figure2}
\end{figure}
This relates to an expanding universe with a scale factor
$a(t)$ given by $a(t)\sim t^p$, where
\begin{eqnarray}
   p \equiv {1\over 3x^2_0 -\sqrt{6}\lambda_1 x_0} = {1\over 3x^2_0 +2\sqrt{6m}x_0}  \ .
   \label{slope:p}
\end{eqnarray}
The linear perturbation about the points $x_{0+}=+1$ and
$x_{0-}=-1$ give the eigenvalues $\mu_{+}=6+4\sqrt{6m}$ and
$\mu_{-}=6-4\sqrt{6m}$, respectively. Thus for positive $m$,
$x_{0+}=+1$ is always unstable and $x_{0-}=-1$ is stable for
$m>3/8$ but unstable for $m<3/8$. Moreover, in the linear
perturbation about the Lorentz violation potential-kinetic
solution, we obtain the eigenvalue $\mu=8m-3$. The solution is
stable for $m<3/8$. In Figs.~\ref{figure1} and \ref{figure2}, we
show the phase plane plot for $m>3/8$ and $m<3/8$. We note that
the trajectories are confined inside the circle given by
$x^2+y^2=1$.

Another remarkable feature of the above model is that the equation of
state is given by
\begin{eqnarray}
   \omega_\phi = -1 + {16\over 3} m  \ ,
\end{eqnarray}
completely determined by the parameter $m$ of the coupling vector.
Thus, we always have $\omega_\phi >-1$ for ordinary scalar field.

%===============================================================%
%************************ SECTION VI ***************************%
%===============================================================%
\section{Conclusions}
In this paper, we have studied the dynamics of a scalar field in the Lorentz violating
scalar-vector-tensor theory of gravity, taking into account the effect of the power-law effective coupling vector. Since the effective coupling vector be dynamics variable, the equation of state is dependent on the coupling parameter. For the model with the power law Hubble parameter and coupling vector, we find an exact solution of the equation of state. A constant equation of state corresponds to $n=2$ while for $n >2$ leads to dynamics equation of states. In this case, the scalar fields are completely associated with the Lorentz violation.

Also, the different form in the coupling vector and the potential models lead to the different qualitative evolution in two regimes of inflation. The results show that, for the inverse power-law potential, the Hubble parameter decreases during the Lorentz violation stage and increases in the standard slow roll stage. For the  power-law potential, the Hubble parameter increases  during the Lorentz violation stage but it decreases in the standard slow roll stage.

From the qualitative study of the dynamical system, we have demonstrated the attractor behavior of inflation driven by a scalar field in the context of scalar-vector-tensor theory of gravity. We have found that there exists the Lorentz violating kinetic dominated solution and the Lorentz violating potential-kinetic dominated solution, depending on the region of the coupling parameter in the simplest Lorentz violating chaotic inflation model. The quadratic coupling vector and the chaotic potential correspond to the constants  $\lambda_1=\lambda_2=-2\sqrt{m}$ and $\Gamma_1=\Gamma_2=1/2$. There are two important results of this study, which are different from the scalar-tensor theory of gravity: the condition for the accelerating universe, Eq.~({\ref{self-sim}}) and the slope, $p$, Eq.~({\ref{slope:p}}). The first one yields $\lambda_1{\bf x}_0 >\sqrt{3/8}(\omega_\phi+1/3)$. The analysis of the critical points show that we may obtain an accelerated expansion provided that the solutions are approaching the Lorentz violation kinetic dominated solution with $m>1/6$ and approaching the Lorentz violation potential-kinetic dominated solution with $m<3/8$. When the accelerating condition is satisfied, the slope $p$ characterizes the properties of the inflating universe: power-law inflation ($p>0$), de Sitter inflation ($p=0$) and superinflation ($p\equiv -|p|<0$). In other cases, if $\lambda_1$ and $\lambda_2$ are constants, one finds that the coupling vector is still quadratic in scalar field, $\bar{\beta}\sim \phi^2$,  while the potential as a function of scalar field $\phi$ is given by  a power-law potential, $V(\phi) \sim \phi^{2\gamma}$ and $\Gamma_1=1/2$, $\Gamma_2=1-1/2\gamma$, where $\gamma = \lambda_2/\lambda_1$. Moreover, in order to obtain dynamical evolution of the system, we
need to solve Eqs.~(\ref{auto-L1}) and (\ref{auto-L2}) together with Eqs.~(\ref{auto-x}) and (\ref{auto-y}). For a realistic model, the effect of an additional component (matter field) would be interesting~\cite{ari}.

Finally, we would like to emphasize that there exists an attractor solution in the Lorentz violating scalar-vector-tensor theory of gravity.

\acknowledgments{Arianto and Supardi would like to thank BPPS,
Dikti, Depdiknas, Republic of Indonesia for financial support. They
also wishe to acknowledge all members of Theoretical Physics
Laboratory, Faculty of Mathematics and Natural Sciences, ITB, for warmest hospitality.
This work is partially supported by Research KK ITB 2007 No.
174/K01.07/PL/07.}

\end{document}